\documentclass{sig-alternate}

\usepackage{graphicx}
\usepackage{amsmath}
\usepackage{amssymb}
\usepackage{color}
\usepackage{ifpdf}

\usepackage{float}
\usepackage[utf8]{inputenc}
\usepackage{multirow}
\usepackage{rotating}
\usepackage{subfigure}
\usepackage{setspace}

\usepackage{moresize}

\usepackage{url}
\usepackage{booktabs}
\usepackage{listings}
\usepackage{paralist}
\usepackage{wrapfig}
\usepackage{multirow}
\usepackage{ifpdf}
\usepackage{xspace}
\usepackage{keyval}
\usepackage{color}

\definecolor{listinggray}{gray}{0.95}
\definecolor{darkgray}{gray}{0.7}
\definecolor{commentgreen}{rgb}{0, 0.4, 0}
\definecolor{darkblue}{rgb}{0, 0, 0.4}
\definecolor{middleblue}{rgb}{0, 0, 0.7}
\definecolor{darkred}{rgb}{0.4, 0, 0}
\definecolor{brown}{rgb}{0.5, 0.5, 0}

\usepackage[normalem]{ulem}
\makeatletter
\def\cyanuwave{\bgroup \markoverwith{\lower3.5\p@\hbox{\sixly \textcolor{cyan}{\char58}}}\ULon}
\def\reduwave{\bgroup \markoverwith{\lower3.5\p@\hbox{\sixly \textcolor{red}{\char58}}}\ULon}
\def\blueuwave{\bgroup \markoverwith{\lower3.5\p@\hbox{\sixly \textcolor{blue}{\char58}}}\ULon}
\font\sixly=lasy6 
\makeatother

\newif\ifdraft
\ifdraft
\usepackage{xcolor}
\definecolor{ocolor}{rgb}{1,0,0.4}
\newcommand{\onote}[1]{ {\textcolor{ocolor} { (***Ole: #1) }}}
\newcommand{\terminology}[1]{ {\textcolor{red} {(Terminology used: \textbf{#1}) }}}

\newcommand{\jhanote}[1]{ {\textcolor{red} { ***shantenu: #1 }}}
\newcommand{\alnote}[1]{ {\textcolor{blue} { ***andreL: #1 }}}
\newcommand{\amnote}[1]{ {\textcolor{blue} { ***andreM: #1 }}}
\newcommand{\smnote}[1]{ {\textcolor{brown} { ***sharath: #1 }}}
\newcommand{\pmnote}[1]{ {\textcolor{brown} { ***Pradeep: #1 }}}
\newcommand{\msnote}[1]{ {\textcolor{cyan} { ***mark: #1 }}}
\newcommand{\mrnote}[1]{ {\textcolor{purple} { ***melissa: #1 }}}
\definecolor{orange}{rgb}{1,.5,0}
\newcommand{\aznote}[1]{ {\textcolor{orange} { ***ashley: #1 }}}
\definecolor{dandelion}{cmyk}{0,0.29,0.84,0}
\newcommand{\mtnote}[1]{ {\textcolor{dandelion} { ***matteo: #1 }}}
\newcommand{\note}[1]{ {\textcolor{magenta} { ***Note: #1 }}}
\else
\newcommand{\onote}[1]{}
\newcommand{\terminology}[1]{}

\newcommand{\alnote}[1]{}
\newcommand{\amnote}[1]{}
\newcommand{\athotanote}[1]{}
\newcommand{\smnote}[1]{}
\newcommand{\pmnote}[1]{}
\newcommand{\jhanote}[1]{}
\newcommand{\msnote}[1]{}
\newcommand{\mrnote}[1]{}
\newcommand{\aznote}[1]{}
\newcommand{\mtnote}[1]{}
\newcommand{\note}[1]{}
\fi

\newcommand{\pilot}{Pilot\xspace}
\newcommand{\pilots}{Pilots\xspace}
\newcommand{\pilotjob}{Pilot-Job\xspace}
\newcommand{\pilotjobs}{Pilot-Jobs\xspace}
\newcommand{\pilotcompute}{Pilot-Compute\xspace}

\newcommand{\pilotdata}{Pilot-Data\xspace}
\newcommand{\pilotdatainmem}{Pilot-Data Memory\xspace}

\newcommand{\computeunit}{Compute-Unit\xspace}
\newcommand{\computeunits}{Compute-Units\xspace}
\newcommand{\dataunit}{Data-Unit\xspace}
\newcommand{\dataunits}{Data-Units\xspace}
\newcommand{\du}{DU\xspace}
\newcommand{\dus}{DUs\xspace}

\newcommand{\cu}{CU\xspace}
\newcommand{\cus}{CUs\xspace}

\lstdefinestyle{myListing}{
  frame=single,
  backgroundcolor=\color{listinggray},
  language=C,
  basicstyle=\ttfamily \footnotesize,
  breakautoindent=true,
  breaklines=true
  tabsize=2,
  captionpos=b,
  aboveskip=0em,
  belowskip=-2em,
}

\lstdefinestyle{myPythonListing}{
  frame=single,
  backgroundcolor=\color{listinggray},
  language=Python,
  basicstyle=\ttfamily \scriptsize,
  breakautoindent=true,
  breaklines=true
  tabsize=2,
  captionpos=b,
}

\ifpdf
\DeclareGraphicsExtensions{.pdf, .jpg, .tif}
\else
\DeclareGraphicsExtensions{.ps,  .eps, .jpg}
\fi

\usepackage{listings}
\usepackage{paralist}
\lstnewenvironment{code}[1][]%
{
\noindent
\minipage{1.0 \linewidth}
\vspace{0.5\baselineskip}
\lstset{
    language=Python,
    frame=single,
    captionpos=b,
    stringstyle=\ttfamily,
    basicstyle=\scriptsize\ttfamily,
    showstringspaces=false,#1}
}
{\endminipage}

\defaultleftmargin{1em}{}{}{}

\begin{document}
\title{Pilot-Abstraction: A Valid Abstraction for Data-Intensive
  Applications on HPC, Hadoop and Cloud Infrastructures?}

\numberofauthors{1} 
\author{\alignauthor Andre Luckow$^{1,2}$, Pradeep Mantha$^{1}$, Shantenu Jha$^{1*}$ \vspace{2mm}\\ 
\affaddr{\emph{ $^{1}$ RADICAL, ECE, Rutgers University, Piscataway, NJ 08854, USA}} \\ \affaddr{\emph{$^{2}$ School of Computing, Clemson University, Clemson, SC 29634, USA}} \\
\email{\emph{$^{(*)}$Contact Author: shantenu.jha@rutgers.edu}}
} 

\date{}
\maketitle

\begin{abstract}
    
  HPC environments have traditionally been designed to meet the
  compute demand of scientific applications and data has only been a
  second order concern. With science moving toward data-driven
  discoveries relying more and more on correlations in data to
  form scientific hypotheses, the limitations of existing HPC
  approaches become apparent: Architectural paradigms such as the
  separation of storage and compute are not optimal for I/O intensive
  workloads (e.\,g. for data preparation, transformation and SQL
  workloads). While there are many powerful computational and
  analytical kernels and libraries available on HPC (e.\,g. for scalable
  linear algebra), they generally lack the usability and variety of
  analytical libraries found in other environments (e.\,g.\ the Apache
  Hadoop ecosystem). Further, there is a lack of abstractions that
  unify access to increasingly heterogeneous infrastructure (HPC,
  Hadoop, clouds) and allow reasoning about performance trade-offs in
  these complex environments.  At the same time, the Hadoop ecosystem
  is evolving rapidly with new frameworks for data processing and has
  established itself as de-facto standard for data-intensive workloads
  in industry and is increasingly used to tackle scientific problems.
  In this paper, we explore paths to interoperability between Hadoop
  and HPC, examine the differences and challenges, such as the
  different architectural paradigms and abstractions, and investigate
  ways to address them.  We propose the
  extension of the \pilot-Abstraction to Hadoop to serve as
  interoperability layer for allocating and managing resources across
  different infrastructures providing a degree of unification in the
  concepts and implementation of resource management across HPC,
  Hadoop and other infrastructures. For this purpose, we integrate
  Hadoop compute and data resources (i.\,e.\ YARN and HDFS) with the 
  \pilot-Abstraction.

  In-memory capabilities have been successfully deployed to enhance
  the performance of large-scale data analytics approaches (e.\,g.\
  iterative machine learning algorithms) for which the ability to
  re-use data across iterations is critical. 
  As memory naturally fits in with the \pilot concept of retaining 
  resources for a set of tasks, we propose the extension of the   
  \pilot-Abstraction to in-memory resources.  These enhancements to the
  \pilot-Abstraction have been implemented in BigJob. Further, we
  validate the abstractions using experiments on cloud and HPC
  infrastructures investigating the performance of the \pilotdata and
  \pilot-Hadoop implementation, HDFS and Lustre for Hadoop MapReduce
  workloads, and \pilotdatainmem for KMeans clustering. Using
  \pilot-Hadoop we evaluate the performance of Stampede, a
  compute-centric resource, and Gordon, a resource designed for
  data-intensive workloads providing additional memory and flash
  storage. Our benchmarks of \pilotdatainmem show a significant
  improvement compared to the file-based \pilotdata for KMeans with a
  measured speedup of 212.
  
\end{abstract}

\section{Introduction}

As more scientific disciplines rely on data as an important means for
scientific discovery, the demand for infrastructures that support
data-intensive tasks in addition to traditional compute-intensive
tasks, such as modeling and simulations, is increasing. For example,
in biology and astronomy scientific discovery is increasingly based on
analysis of data collected from machines, such as genome sequencing
machines or observatories~\cite{national2014Future}.  Across
disciplines there is a move towards data-driven discovery and with
increasing diversity in the source of data (c.f. the Internet of Things
and the usage of networked sensors to collect data). The term ``fourth
paradigm''~\cite{hey2009} refers to scientific discovery based on data
in addition to theory, experimentation and simulation based discovery.

Data-intensive applications are associated with a wide variety of
characteristics and properties, as summarized by Fox et
al.~\cite{bigdata-ogres,bigdata-use-cases-nist}. Often, they are more
complex and heterogeneous than HPC applications as they typically
comprise of multiple stages with different characteristics. Typical
stages are: data ingest, pre-processing, feature-extraction and
advanced analytics. While some of these stages are I/O bound with
potential different I/O characteristics (random vs.  sequential
access), some stages (e.\,g.\ advanced analytics) are compute- and
memory bound. Managing and supporting such heterogeneous application
workflows on top of heterogeneous resources at scale represents an
important challenge.

HPC infrastructures introduced parallel filesystems, such as Lustre or
GPFS, to meet the increased I/O demands of data-intensive applications
and archival storage to address the need for retaining large volumes
of primary simulation output data. The parallel filesystem model of
using large, optimized storage clusters exposing a POSIX compliant
rich interface and connecting it to compute nodes via a fast
interconnects works well for compute-bound task. It has however, some
limitations for data-intensive, I/O-bound workloads that require a
high sequential read/write performance. Similarly, high-throughput
infrastructures (HTC), such as OSG, rely on separate storage
environments, e.\,g.\ SRM or iRODS based, to manage data. In clouds,
object stores are a common mechanism for persisting data outside of
ephemeral VM-based compute resources. To address some of these issues,
new -- mostly hardware -- capabilities have been added to HPC
resources to meet the demands of data-intensive applications; machines
such as Gordon or Wrangler, provide large memory nodes to facilitate
shared memory data analytics tools (e.\,g.\ R, Python) and additional
storage tiers (e.\,g. SSD).

Hadoop~\cite{hadoop} and the MapReduce abstraction~\cite{mapreduce}
were developed with data as first order consideration and established
the the de-facto standard for data-intensive computing. The biggest
differentiator of Hadoop compared to HPC systems is data-locality:
while HPC systems generally rely on fast interconnects between compute
and storage, Hadoop co-locates compute and data. The Hadoop ecosystems
-- in the following referred to as \emph{Apache Big Data Stack (ABDS)}
-- provides a manifold set of novel tools and higher-level
abstractions for data processing. In addition to MapReduce,
Spark~\cite{Zaharia:2010:SCC:1863103.1863113} gained popularity for
memory-centric data processing and analytics.
ABDS tools and frameworks are increasingly used in sciences
(see~\cite{Dede:2011:REM:2132876.2132888,Massie:EECS-2013-207}). While
the Hadoop platform has proven its value in scientific applications,
challenges remain in deploying ABDS applications on HPC infrastructure
and when integrating these with HPC applications, e.\,g.\ based on
MPI~\cite{doe_hpcor,nrc_future_2017_2020}.

A main differentiator of ABDS are high-level abstractions that trade-off
capabilities and performance; MapReduce and Spark's RDDs offer the ability to
reason about data processing steps (e.\,g.\ filtering, transformations and
aggregations) without the need to explicitly implement data parallelism.
Further important capabilities are offered in the advanced analytics and
machine learning domain. MPI provides a good abstraction for implementing
parallel applications providing primitives for point-to-point and collective
communications; however, it lacks the productivity of higher-level ABDS
abstractions. 

Over the past decades, the \emph{High Performance Distributed
  Computing (HPDC)} community has made significant advances in
addressing resource and workload management on heterogeneous
resources. In contrast, the ABDS ecosystem has been evolving to
provide similar levels of sophistication for commercial enterprise.
In fact, some conceptual ideas dominant in the HPDC world have already
started making an impact in ABDS tools. For example, the concept of
multi-level scheduling as manifest in the decoupling workload
assignment from resource management via the concept of intermediate
container jobs (also referred to as \pilotjobs~\cite{pstar12}) has
made its presence in ABDS after the transition from Hadoop to
YARN/Mesos. This concept has been adapted and refined in the ABDS
environment allowing frameworks to retain resources (cores, memory,
storage) and expand/shrink the resource pool if necessary.  It turns
out to be the case that multi-level scheduling is even more important
for data-intensive applications as often only application-level
schedulers can be aware of the localities of the data sources used by
a specific application.  This motivated the extension of the
\pilot-Abstraction to support data-aware scheduling on
application-level~\cite{pilotdata}. As most HPC schedulers are data
agnostic, this is an important/critical extension of capabilities.
However, this advance has had the interesting consequence that
heterogeneity in ABDS systems is now no longer confined to the
filesystems and resource-access mechanisms, but like traditional HPDC
systems, the resource management interface and semantics are
different.

Collectively, the above features and trends point to the possibility
and need for consilience between the HPDC and ABDS approaches for
resource management. The aim of this paper is to examine the need for
consilience in resource management approaches between ABDS and
traditional HPDC approaches, understand some of the challenges on the
path to doing so and explore the \pilot-Abstraction as one possible
way to address some of the challenges.

Another concern of this paper is to understand how to address the issues
of interoperability: There is a great need to integrate both HPDC and
Hadoop, e.\,g.\ due to the necessity to co-locate data/compute or to
combine compute-centric HPC applications (e.\,g.\ linear algebra
solvers) with ABDS applications (e.\,g.\ MapReduce and Spark). In this
paper, we explore the usage of the \pilot-Abstraction inside an ABDS
environment, as well as a path for the interoperable use of ABDS
applications on HPDC infrastructure.

This paper makes the following contributions: (i) We propose several
extensions to the \pilot-Abstraction~\cite{pstar12} to better
facilitate data processing and advanced analytics on HPDC and to
support interoperability with Hadoop; specifically, we design and
implement \pilot-Abstractions that provides a common approach for
data-aware resource management on and across HPC, cloud and Hadoop
infrastructures. By supporting Hadoop's resource manager YARN, the
\pilot-Abstraction can be used as standard application framework
simplifying the usage of HPC application on Hadoop. (ii) We extend an
implementation of the \pilot-Abstraction to facilitate the deployment
and execution of ABDS applications on HPC. \pilot-Hadoop enables the
dynamic, ad-hoc creation of Hadoop or Spark clusters on HPC
infrastructures. (iii) We extend \pilotdata for distributed in-memory
computing that is essential for scalable analytics, such as iterative
machine learning. \pilotdatainmem provides a unified way to access
distributed memory within data-intensive applications that is
integrated with the data-affinity model of \pilotdata. \pilotdata
offers a unified approach for data management across complex storage
hierarchies comprising of local disks, cloud storage, parallel
filesystems, SSD and memory. (iv) Finally, we validate the proposed
abstractions and tools on XSEDE using Stampede and Gordon (machine
designed for data-intensive applications). We investigate the
performance of HDFS and Lustre using a MapReduce workload and of
KMeans running on different ABDS and HPC-backends using the
\pilot-Abstraction.        
      
This paper is structured as follows: In section~\ref{sec:related} we
survey the ABDS ecosystem and compare the provided tools and
abstractions with the HPC environment.
We continue with a discussion of the new \pilotdata capabilities that support
Hadoop/HPC interoperability as well as advanced analytics and machine learning
applications in section~\ref{sec:pilot-data-hadoop}. The results of our
experimental validation are presented in section~\ref{sec:experiments}. We
conclude with a discussion of the contributions and lessons learn as well as
relevant future issues in section~\ref{sec:conclusion}.

\section{Background and Related Work}
\label{sec:related}

Hadoop emerged separate from high-performance computing in enterprise
environments inspired by Google's cloud infrastructure based on the Google
Filesystem and MapReduce~\cite{mapreduce}. Apache Hadoop evolved to a general
purpose cluster computing framework suited for many kinds of data-intensive
applications~\cite{tale-two-architectures}. While Hadoop MapReduce lacks behind
some capabilities, e.\,g., high-performance inter-process communication, a set
of novel, high-level abstractions and runtime environments emerged on top of
the core Hadoop services, the Hadoop Filesystem (HDFS) and the YARN resource
manager. YARN allows the deployment of any application on a Hadoop cluster 
without the need to retrofit applications into the MapReduce model as required 
for Hadoop 1.

In this section, we summarize our previous work in analyzing the
characteristics and properties HPC and Hadoop Infrastructures as well as
related work to provide abstractions and runtime environment for such
applications.

\subsection{HPC and ABDS Abstractions}

This section describes abstractions and runtime systems for data-intensive
computing in the HPC and ABDS domains. As alluded to earlier, data-intensive
applications are typically more heterogeneous than compute-intensive
simulations. Data-intensive applications are often modeled as a pipeline of
tasks (from data ingest, storage, processing to analytics) or a direct acyclic
graph with tasks corresponding to nodes in the graph. We use the term
\emph{workflow} to refer to multi-stage data-processing; \emph{data pipeline}
is often used as a synonym for this. 

A common concern is the provision of a scalable environment support different
stages of data processing and analytics workflows: from coarse-grained data
parallelism for data-filtering with MapReduce to fine-grained parallelism for
machine learning that often relies on scalable linear algebra libraries.
In this section, we give a brief overview of HPC libraries and abstractions and
investigate their usage within data-intensive workflows. We further compare and
contrast these approaches to abstractions developed in the ABDS world.

\subsubsection*{Communication-Abstractions (MPI)}

Analytics and machine learning algorithms can often be expressed and
transformed into matrix operations and thus, require efficient linear algebra
implementation. MPI and OpenMP are the standard abstraction for implementing
parallel applications on HPC infrastructures. While they provide an important
building block for scalable analytics, they lack the usability of higher-level
abstractions (such as the R dataframe). Several parallel, numerical libraries
(MPI/OpenMP-based) that support analytics have been developed, e.\,g.\
ScaLAPACK~\cite{scalapack} or ARPACK~\cite{arpack}. These parallel libraries
are based on the lower-level libraries LAPACK and BLAS. While they provide an
important building block for scalable analytics, they lack the usability of
higher-level abstractions. Which the introduction of
YARN, the execution of MPI applications on Hadoop clusters is well supported,
e.\,g.\ MPICH2-YARN and Hamster, an YARN application master for OpenMPI.

\subsubsection*{File-based Abstractions}

Scientific applications are commonly based on files. Many abstractions for
improving file management in the HPC context emerged: Filecules~\cite{1652137}
e.\,g.\ simplify the the processing of file groups. Similarly, other data
management systems, such as iRods often work on collection/groups of files.
However, file-based data management is often associated with inefficiency in
particular for temporary data, e.\,g.\ intermediate data of iterative machine
learning applications. iRods~\cite{Rajasekar:2010:IPI:1855046} utilizes
so-called collections to group and manage files. While these tools allow a
logical grouping of data, they provide limited support for data-parallel
processing of these files as data partitioning and processing is done outside
of these tools.

\subsubsection*{\pilotjobs and Workflows}

\pilotjobs~\cite{pstar12} have been developed to support ensembles of
fine-grained tasks in HPC environments; examples of \pilotjobs are
Condor-G/GlideIn~\cite{glidein} or BigJob~\cite{saga_bigjob_condor_cloud}. By
using a set of placeholder jobs distributed across multiple resources, the
system can accommodate dynamic task workloads through a dynamically adjusting
resource pool improving the overall utilization at the same time. \pilotjobs
have been successfully used to enhance the performance of scientific workflows.

Scientific workflows popularized the direct acyclic graphs (DAGs) abstraction
as fundamental way to express workflows; examples of such systems are
Pegasus~\cite{Deelman-FGCS-2014} and Taverna~\cite{Wolstencroft01072013}. Many
of these system focus on the higher-level workflow abstraction lacking a
high-performance, vertically integrated runtime system (as provided by the ABDS
stack). Another constraint is the fact that the dataflow is often based on
files. In ABDS similar DAG abstractions emerged, which are implemented on top
of ABDS primitives, such as MapReduce, and are discussed in the following
section.

\pilotjobs have been extended to facilitate data management (mainly on
file-basis) and for dataflow-oriented workflows. Falcon~\cite{1362680} provides
a data-aware scheduler on top of a pool of dynamically acquired compute and
data resources~\cite{Raicu:2008:ALD:1383519.1383521}. The so called data
diffusion mechanism can cache data on \pilot-level enabling the
efficient re-use of data. 
Another area of research is the utilization of distributed memory for
data-intensive, task-based workflows. Swift/T~\cite{10.1109/CCGrid.2013.99} is
a rewrite of the Swift to utilize the MPI-based Turbine engine for processing
of data-intensive workflows benefiting from MPI features, such as effective
collective communications.

\subsubsection*{MapReduce \& Higher-Level Abstractions} 

MapReduce~\cite{mapreduce} proofed an effective abstraction for processing data
in a parallel way decoupling storage backend, data formats and processing
engine. Hadoop MapReduce established itself as de-facto standard for scalable
data processing -- in contrast to file-based approaches found in the scientific
workflow community -- MapReduce hides complex details, such as data
organization, formats, data partitioning and aggregation. While MapReduce
simplified the creation of data-intensive application (particularly
applications that need to process vast volumes of data), the MapReduce
abstraction is limited in its expressiveness as pointed out by various
authors~\cite{magalan,dryad} and lead to manifold higher-level abstractions for
implementing sophisticated data pipelines.

In addition, the native APIs provided by Hadoop are complex: the creation of a
simple application requires an implementation of a map and reduce function, as
well as various configurations and auxiliary functions. Also, the creation of
more complex data pipelines or iterative machine learning applications
consisting of multiple MapReduce jobs is very complex. Thus, a set of
high-level APIs, such as Apache Crunch~\cite{crunch},
Cascading~\cite{cascading}, Apache Flink~\cite{flink} and Spring
XD~\cite{spring_xd}, emerged. While the expressiveness of the API was better
than MapReduce, they still were constrained by the MapReduce runtime system.

With the emergence of YARN, several new processing engines emerged in the
Hadoop ecosystems that improved the support for workflows. Framework, such as
Spark~\cite{Zaharia:2010:SCC:1863103.1863113} and Tez~\cite{tez}, provide
richer abstractions that are built on modern processing engines that can retain
resources across task generations and effectively utilize distributed memory.
Spark's reliable distributed dataset (RDD) abstraction provides a powerful way
to manipulate distributed collection stored in the memory of the cluster nodes.
Spark is increasingly used for building complex data workflows and advanced
analytic tools, such as MLLib~\cite{mllib} and SparkR~\cite{spark-r}.

As alluded, the ability to utilize efficient collectives is important for
advanced analytics implementations that often require fast linear algebra
implementations. Some hybrid frameworks have been proposed to provide MPI-style
collective operations in conjunction with data-intensive abstractions, such as
MapReduce. For example, Harp~\cite{harp} proposes the usage of collective
operations in the map-phase of a MapReduce application -- the model is referred
to Map-Collective. This model is similar to the bulk-synchronous communication
model or the MPI communication model; Harp however aims to provide a
higher-level abstraction than MPI. In contrast to other abstraction, such as
MapReduce, the user is required to manually manage data partitions and
processes. Different implementations and algorithms for the HARP collective
layer have been investigated, e.\,g.\ based on based e.g. on the Netty and
Azure inter-role communication mechanism. A constraint is that the collective
framework of Harp currently only supports a static group of resources.

\subsubsection*{Dataset and Dataframe Abstractions} 

The \emph{dataset} or \emph{dataframe} abstractions originally introduced in R
(respectively his predecessor S) exposes data in a tabulated format to the
user and supports the efficient expression of data transformations and
analytics~\cite{r-lang}. A dataframe typically stores data matrix of different
types of data (e.\,g.\ numeric, text data and categorical data). The dataframe
abstraction supports various functions for manipulating data stored inside the
data structures, e.\,g.\ to subset, merge, filter and aggregate data, using
well-defined primitives or SQL. Similar abstractions emerged for other
languages and runtime environments, e.\,g.\ Pandas~\cite{pandas},
Scikit-Learn~\cite{DBLP:journals/corr/BuitinckLBPMGNPGGLVJHV13} and Dato
SFrame~\cite{sframe} for Python. For Mahout a dataframe abstraction has been
proposed.

While the dataframe abstractions are very expressiveness and well-suited for
suited for implementing advanced analytics, they traditionally lacked the
ability to scale-out. The MLI-inspired Spark Pipeline API~\cite{mli} for Spark,
H2O~\cite{h2o}, and Blaze~\cite{blaze} for Python attempt to introduce similar
abstractions also supporting scalable backends - however, they currently lack
in the variety of analytics frameworks and algorithms available for these
tools. Another constraint is that many of these framework, e.\,g.\ Spark, are
Java-based making it difficult to integrate these with native libraries,
e.\,g.\ parallel linear algebra libraries from the scientific computing domain.
HPC abstractions in contrast to the dataframe or MapReduce abstraction focus on
low-level functions (e.\,g.\ MPI for communications). Thus, they often do not
provide an easy way to explore data-parallelism. Further, higher-level
dataset/dataframe abstractions designed for domain scientists are typically not
available.

The ability to combine different frameworks and abstractions is the key for
implementing end-to-end data-intensive workflows with multiple stages with
different workload characteristics. To address these complex requirements, it
can be expected, that the design space for abstractions will be further
explored and more hybrid approaches will emerge allowing for efficiently
supporting end-to-end data pipelines (including analytics). Another constraint
is the fact, that the majority of the available runtime systems is based on
Java. Python -- a popular language for scientific computing -- is
typically only supported as secondary option.

\subsection{HPC and Hadoop}

The heterogeneity of distributed infrastructures is still increasing: HPC and
Hadoop e.\,g.\ following dramatically different design paradigms. In HPC
environments traditionally storage and compute are separated connected by an
high-end network (e.\,g.\ Infiniband). To address the increasing need to
process large volumes of data on HPC, the capacity of these storage systems and
the networks. System, such as Wrangler~\cite{wrangler}, maintain this
separation. Wrangler in particular deploys a mix of a Lustre storage system, a
SSD-based high-end storage system and local storage. In addition archival
storage systems often based on HPSS are used to store cold data.

\emph{Resource Management:} The fine-grained data-parallelism of data-intensive
applications is ideally suited for multi-level scheduling. Multi-level
scheduling~\cite{1392910} originated in HPC environments and proofed as
efficient mechanism to managed ensembles of tasks. By decoupling system-level
and application-level scheduling and by taking into account system-level, such
as resources utilization and allocation policies, as well as application-level
objectives, e.\,g.\ dynamic resource requirements for every application stage,
in most cases a superior performance can be achieved. \pilotjobs~\cite{pstar12}
provided a powerful abstraction for implementing multi-level scheduling on HPC
systems. While Hadoop originally only provided a rudimentary scheduling system,
the new YARN scheduler provides efficient support for application-level
scheduling.

YARN address the need that with the uptake of Hadoop, the requirements with
respect to resource management increased: more complex data localities (memory,
SSDs, disk, rack, datacenter), long-lived services, periodic jobs, interactive
and batch jobs need to be supported on the same environment. Multi-level
scheduling serves as the basic architectural principle to support this
requirement. YARN~\cite{yarn-paper}, Hadoop's resource, aims to address these
limitations. In contrast, to traditional batch schedulers, YARN is optimized
for data-intensive environments supporting data-locality and the management of
a large number of fine-granular tasks (found in data-parallel applications).
YARN enables applications to deploy their own application-level scheduling
routines on top of Hadoop-managed storage and compute resources. While YARN
manages the lower resources, the higher-level runtimes typically use an
application-level scheduler to optimize resource usage for the application.
Applications need to initialize their so-called Application-Master via YARN;
the Application Master is then responsible for allocating resources --
containers -- for the applications and to execute tasks in these containers.
Data locality, e.\,g.\ between HDFS blocks and container location need to
manually managed by the application master (by requesting containers on
specific nodes/racks etc.). Other resource management systems addressing
similar needs emerged, e.\,g.\ Mesos~\cite{Hindman:2011:MPF:1972457.1972488}
Omega~\cite{omega} Google's distributed scheduler.

Managing resources on top of YARNs is associated with several
challenge: while the default design particularly facilitates the
fine-grained data parallelism of MapReduce, for certain application
characteristics it is desirable to retain resources during longer
period, e.\,g.\ to longer cache data that is often re-used in-memory,
to have readily available resources for interactive applications or to
facilitate iterative processing. Several higher-level frameworks for
YARN addressing specific application characteristics emerged:
Llama~\cite{llama} offers a long-running application master for YARN
designed for the Impala SQL engine. Apache Slider~\cite{apache-slider}
supports long-running distributed application on YARN with dynamic
resource needs allowing applications to scale to additional containers
on demand.  TEZ~\cite{tez} is a DAG processing engine primarily
designed to support the Hive SQL engine allowing the application to
hold containers across multiple phases of the DAG execution without
the need to de-/re-allocate resources.
REEF~\cite{Chun:2013:RRE:2536274.2536318} is a similar runtime
environment that provides applications a higher-level abstractions to
YARN resources allowing it to retain memory and cores supporting
heterogeneous workloads. REEF Retainers can re-use JVM and store data
in the JVM's memory.

\emph{Interoperability:} To achieve interoperability, several frameworks
explore the usage of Hadoop on HPC resources. Resource
managers, such as Condor and SLURM, provide Hadoop support. Further, various
third-party systems, such as SAGA-Hadoop~\cite{saga-hadoop},
JUMMP~\cite{6702650} or MyHadoop~\cite{Krishnan04myhadoop}, exist. A main
disadvantage with this approach is the loss of data-locality, which the
system-level scheduler is typically not aware of. In addition there are several
other limitations associate with that approach: e.\,g.\ the necessity to load
data into HDFS and the ability to achieve higher cluster utilization by more
fine-grained resource sharing.

For Hadoop deployments typically local storage is preferred; nevertheless some
deployments use a HPC-style separation of compute and storage systems. Hadoop
workloads on these HPC systems is supported via a special client library, which
improves the interoperability with Hadoop; it limits however data locality and
the ability for the application to optimize for data placements since
applications are commonly not aware of the complex storage hierarchy. Another
interesting opportunity is the usage of Hadoop as active archival storage -- in
particular, the newly added HDFS heterogeneous storage support is suitable for
supporting this use case.\\
\\

\emph{Summary:} Understanding performance in a heterogeneous, distributed
environments is complex. In the remainder of the paper, we investigate the
\pilot-Abstraction as unifying concept to efficiently support the
interoperability between HPC and Hadoop. By utilizing the multi-level
scheduling capabilities of YARN, \pilotdata can efficiently manage Hadoop
cluster resources providing the application with the necessary means to reason
about data and compute resources and allocation. On the other side, we show,
how the \pilot-Abstraction can be used to manage ABDS application on HPC
environments.

\section{Pilot-Abstraction for Hadoop: Design and Implementation}
\label{sec:pilot-data-hadoop}

The \pilot-Abstraction~\cite{pstar12} has been successfully used in HPDC for
supporting a diverse set of task-based workloads on distributed resources. A
\pilotjob provides the ability to utilize a
placeholder job as a container for a dynamically determined set of
compute tasks. The \pilotdata abstraction~\cite{pilotdata} extends the
\pilot-Abstraction for supporting the management of data in
conjunction with compute tasks. The \pilot-Abstraction defines the
following entities: A \pilotcompute allocates a set of computational
resources (e.\,g.\,cores); a \pilotdata represents space on a physical
storage resource. Further, the abstraction defines a \computeunit (CU)
as a self-contained piece of work represented as executable that is
submitted to the \pilotjob. A \dataunit (DU) represents a
self-contained, related set of data.

\begin{figure}[t]
    \centering
    \includegraphics[width=.5\textwidth]{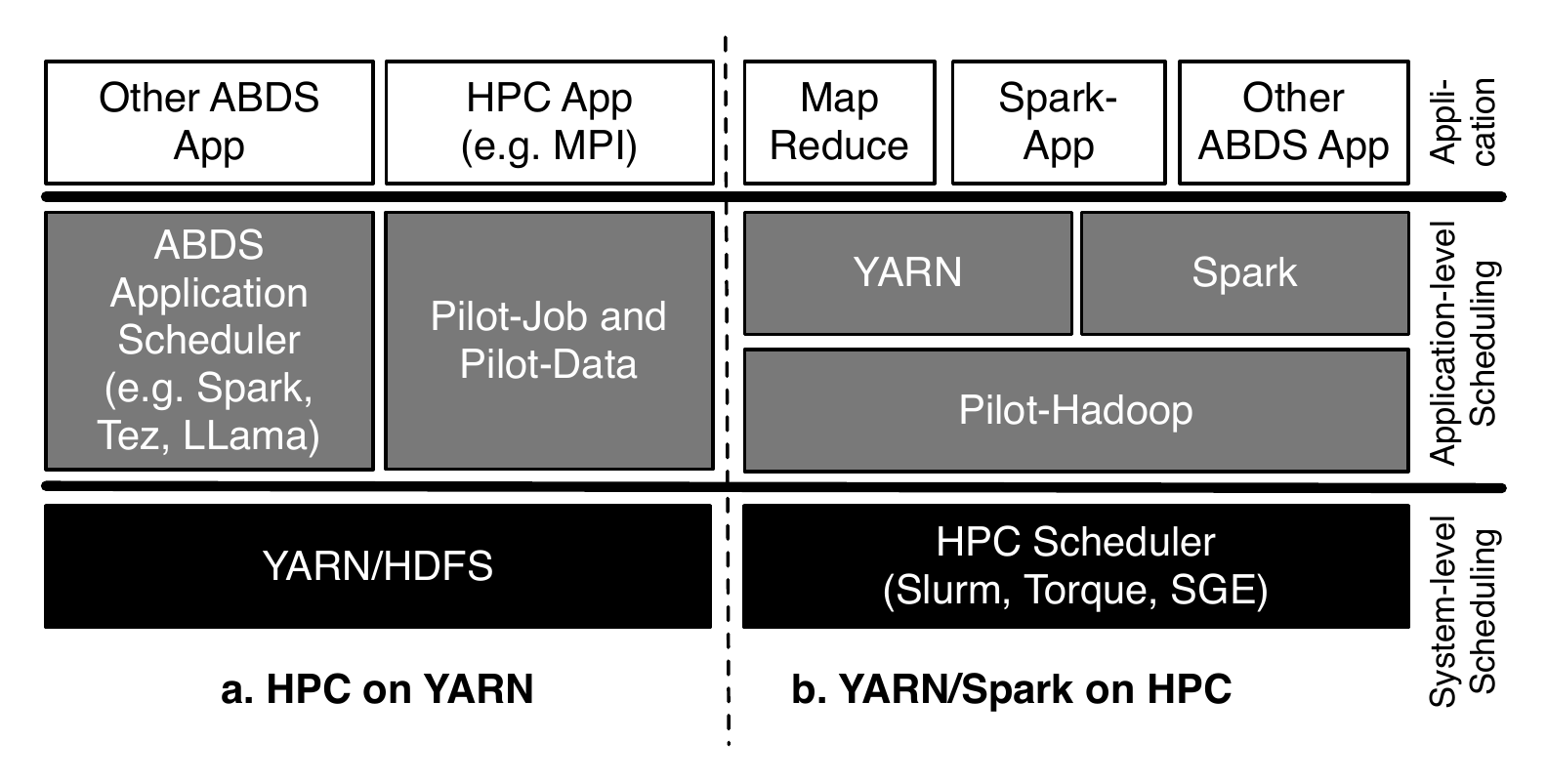}
    \caption{\textbf{Hadoop and HPC Interoperability:} There are two options: 
    a. Running HPC applications inside a YARN 
    cluster, b. spawning a YARN or Spark cluster on a HPC 
    environment.
    \label{fig:figures_hadoop-on-hpc-viceverse}}
\end{figure}

The \pilotjob and \pilotdata abstractions have been implemented within
BigJob~\cite{pstar12,saga_bigjob_condor_cloud}, an interoperable
\pilot-Abstraction framework for supporting heterogeneous tasks-based workloads
on heterogeneous infrastructures. In this section, we present the extensions
made to the \pilot-Abstraction to facilitate a broader set of data-intensive
applications and infrastructures, such as Hadoop. We explore several options
for integrating HPC and Hadoop environments as depicted in
Figure~\ref{fig:figures_hadoop-on-hpc-viceverse} using the \pilot-Abstraction.
By providing support for Hadoop inside BigJob
(Figure~\ref{fig:figures_hadoop-on-hpc-viceverse}a) HPC applications can run
on Hadoop YARN clusters without modification. \pilot-Hadoop enables the
execution of ABDS applications inside environments managed by traditional HPC
schedulers (Figure~\ref{fig:figures_hadoop-on-hpc-viceverse}b).

In section~\ref{sec:interop}, we describe how the \pilot-Abstraction and the
BigJob implementation was extended to support Hadoop resource manager, i.\,e.\
YARN and Mesos, and HDFS storage. We continue with a discussion of
\pilot-Hadoop for running YARN as application-level scheduler on HPC
resources in section~\ref{sec:pilot_hadoop}. Finally, we present
\pilotdatainmem - an infrastructure-agnostic in-memory runtime for analytics
applications in section~\ref{sec:pilotinmem}.

\subsection{Pilot-Abstraction: Interoperable Access and Management of Hadoop Resources}
\label{sec:interop}

With the introduction of YARN, arbitrary applications can be executed within
Hadoop clusters. Nevertheless, utilizing an Hadoop environment outside of
higher-level frameworks, such as MapReduce and Spark, is a difficult task.
Established abstractions that enable the user to reason about compute and data
resources across infrastructures (i.\,e.\ Hadoop, HPC and clouds) are missing.
Also, the new generation of schedulers that emerged in the YARN space impose a
more stringent requirements on the application. While schedulers such as YARN
or Mesos effectively facilitate application-level scheduling, the development
efforts for YARN and Mesos applications are very high. YARN provides e.\,g.\
only a very low-level abstraction for resource management -- the application
must be able to work with a subset of the requested resources. Also, allocated
resources (the so called YARN containers) can be preempted by the scheduler.
Data/compute locality needs to be manually managed by the application scheduler
by requesting resources at the location of an file chunk. To address this,
various frameworks that aid the development of such applications have been
proposed, e.\,g.\ Apache Slider~\cite{apache-slider} and Spring
YARN~\cite{spring-yarn}. While these frameworks simplify development, they do
not address concerns such as interoperability and support for geographically
distributed resources.

To provide a unified, infrastructure-agnostic abstraction, the Hadoop resource
model as exposed by YARN and HDFS must be mapped to the \pilot-Abstraction. The
Hadoop resource model primarily relies on cores and memory for modeling compute
and storage space modeling data resources. While HPC resources typically only
allocate compute cores and nodes, the memory requirement can be easily
translated to a \pilotcompute description of a YARN-based \pilot. Similarly 
HDFS space can be mapped to the space parameter of the \pilotdata description.

\begin{figure}[t]
    \centering
    \includegraphics[width=.51\textwidth]{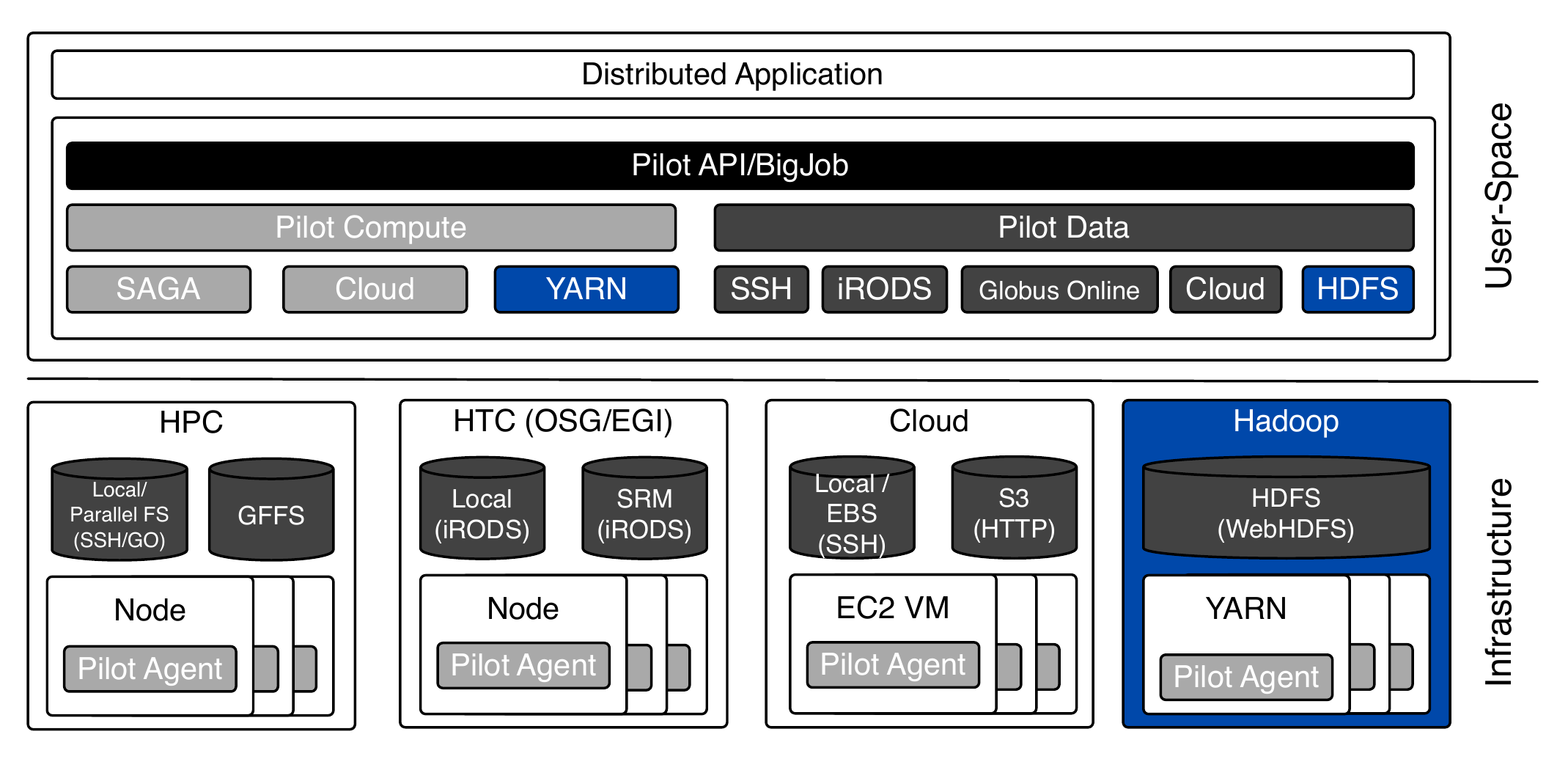}
    \caption{\pilot-Framework Big Job on Heterogeneous Infrastructure: BigJob 
    provide a unified abstraction for compute and data resources. The extension 
    for Hadoop are encapsulated in a YARN adaptor for \pilotcompute and HDFS 
    adaptor for \pilotdata.}
    \label{fig:figures_interoperable_pilot_job}
\end{figure}

To facilitate physical access to Hadoop resources, several adaptors had to be
developed inside the \pilot-Framework BigJob. As shown in
Figure~\ref{fig:figures_interoperable_pilot_job}, the \pilot-Framework is able
to spawn and manage \computeunits and \dataunits on different kinds of compute
and storage resources using a set of adaptors. The new YARN, Mesos and HDFS
adaptors enables applications to take advantage of new infrastructures and
manage their \cus and \dus on dynamic resource pools located in an Hadoop
environment. A particular challenge for the implementation of the YARN adaptor
is the multi-step resource allocation process imposed by YARN, which as alluded
differs significantly from HPC schedulers. The YARN adaptor for BigJob
implements a so-called YARN Application Master, which is the central instance
for managing the resource demands of the application. Once the Application
Master is started, subsequent resource requests are handled by it. The BigJob
Application Master will then request the specified number of YARN container.
Once these are allocated by YARN, the \pilot-Agent will be started inside these
containers. The dispatching of the \cus is done via the normal \pilot-Framework
internal mechanisms without involvement of YARN.

\begin{figure}[t]
    \centering
\includegraphics[width=.43\textwidth]{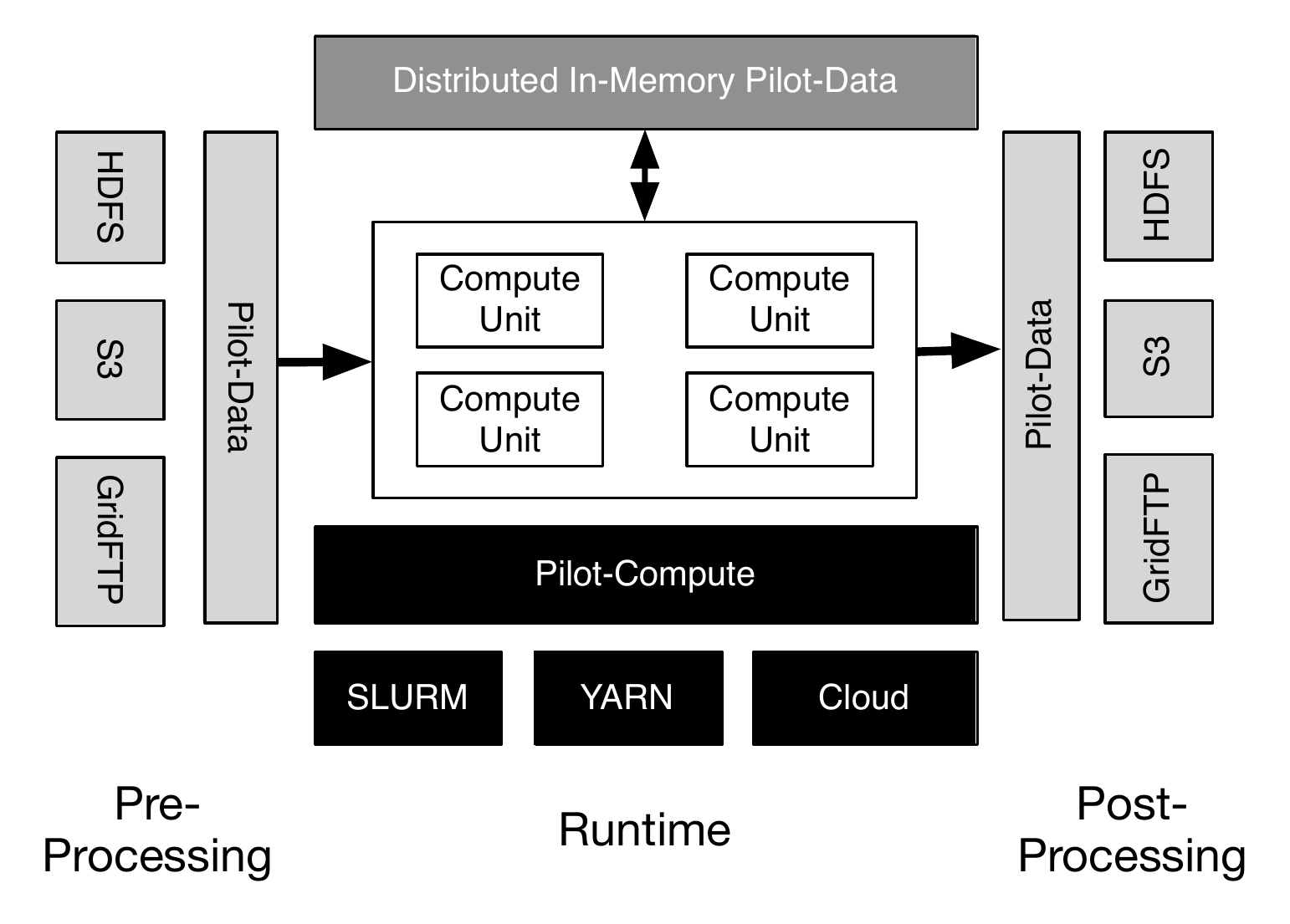}
    \caption{\textbf{Pilot-Abstraction for Data-Parallel Processing:} 
    Pilot-Compute allows the flexible allocation of resources across 
    heterogeneous infrastructures, Pilot-Data enables Compute Units to seamless 
    access datasets on different storage infrastructures. The 
    In-Memory \dataunit facilitates data-parallel processing using a group of 
    \computeunits.}
    \label{fig:figures_pilotabstraction-for-dataparallel-processing}
\end{figure}

The HDFS adaptor access the Hadoop Filesystem using the WebHDFS API. \pilotdata
supports access to data from different sources: data may reside on a local or
mounted shared storage system (e.\,g.\ Lustre storage), HDFS, iRods or another
cloud object store. \pilotdata will ensure that the data will be available
before the \computeunit is started. As shown in
Figure~\ref{fig:figures_pilotabstraction-for-dataparallel-processing} the
\pilot-Abstraction enables the implementation of complex data workflows,
i.\,e.\ the stage-in/out and processing of data residing in different sources.
In section~\ref{sec:pilotinmem}, we discuss the usage of \pilotdata for caching
data during complex processing steps. \pilotdata is data format agnostic,
i.\,e.\ the implementation of access to the data structure is done on
application-level (in the \cu). It supports standard formats, such as text
files, CSV but also advanced columnar formats (bcolz) or HDF5.

The \pilot-API provides a unified API across heterogeneous resources and gives
application-level control to storage and compute resources. The new adaptors
enable applications to seamlessly utilize Hadoop resources inside their
data-intensive workflows. Depending on the application requirements the API is
suited for implementing complex data-intensive workflows as well as running
scalable analytics kernels (optimized for complex storage and memory
hierarchies) on the data. The API relies on affinity labels to manage the
co-location of data and compute (see~\cite{pilotdata}). Using the API
developers can model complex storage hierarchies consisting of archival
storage, cold data storage for raw data, warm and hot storage for pre-processed
data used in the model fitting phase and memory for intermediate results of
iterative machine learning algorithms. The runtime system will ensure that
data/compute will be co-located if possible to improve performance.

\subsection{Pilot-Hadoop: Supporting Application-Level Interoperability for ABDS and HPC}
\label{sec:pilot_hadoop}

Having discussed the extensions to the \pilot-Abstraction to support Hadoop
resources, we explore the usage of Hadoop and other ABDS frameworks on HPC
resources in this section (see Figure~\ref{fig:figures_hadoop-on-hpc-viceverse}
(b)). \pilot-Hadoop~\cite{pilot-hadoop} provides a framework for executing ABDS
applications written for YARN (e.\,g.\ MapReduce) or Spark on HPC and cloud
resources. Pilot-Hadoop has been developed as successor to
SAGA-Hadoop~\cite{saga-hadoop} utilizing the handling of multi-node cluster
environments of the \pilot-Agent.

\begin{figure}[t]
    \centering
\includegraphics[width=.41\textwidth]{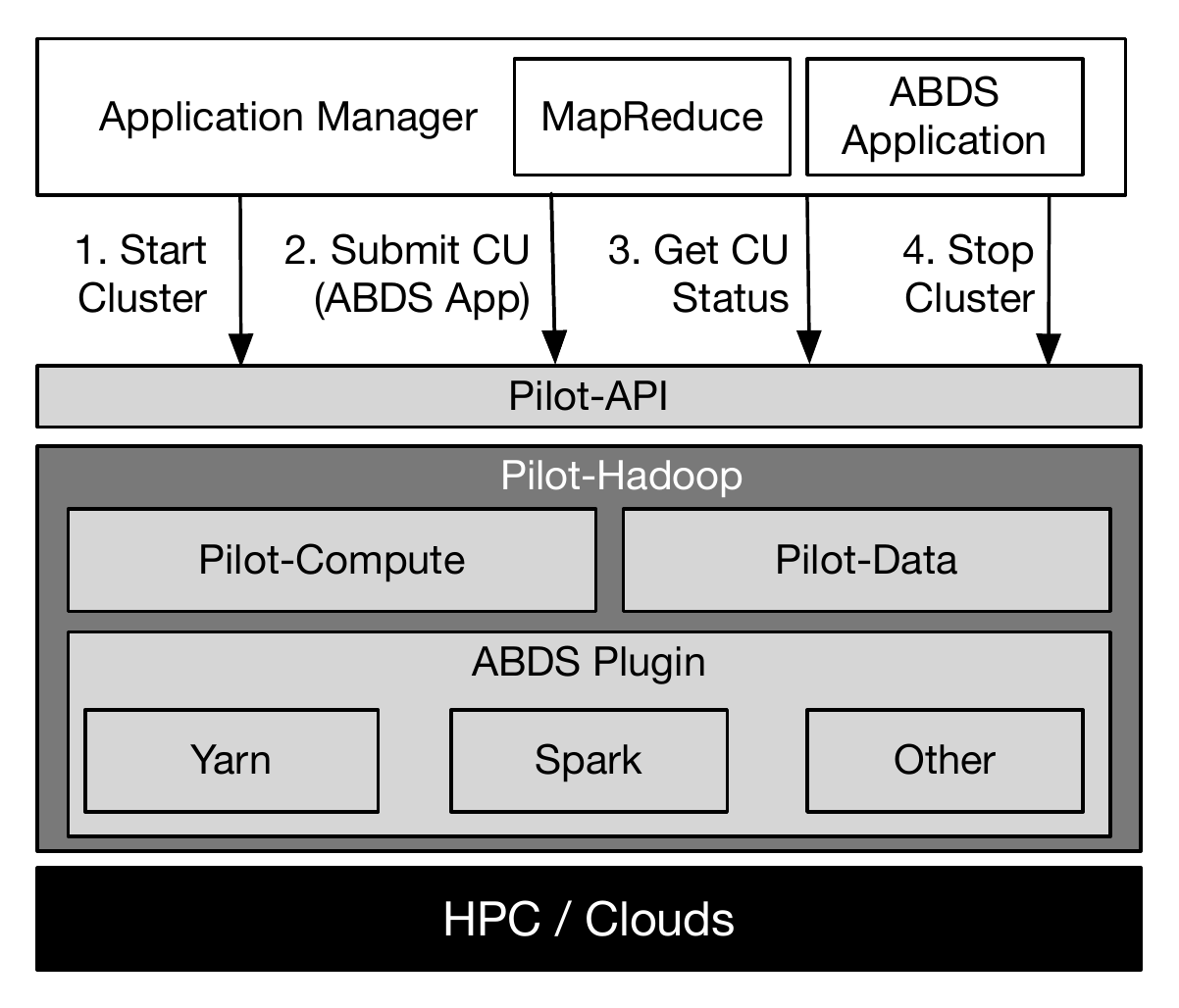}
    \caption{\textbf{Pilot-Hadoop for HPC and Cloud Infrastructures:} 
    \pilot-Hadoop provides  uniform framework to managing Hadoop and Spark 
    clusters on resources managed by HPC schedulers, such as PBS, SGE and 
    SLURM. The \pilot-API provides a unifying interface for allocating a 
    YARN/Spark cluster on HPC and for managing \computeunits that marshal a 
    YARN, Spark or other ABDS application.}
    \label{fig:figures_pilot_abds}
\end{figure}

Figure~\ref{fig:figures_pilot_abds} illustrates the architecture of
\pilot-Hadoop. Pilot-Hadoop uses the \pilot-Abstraction and the
BigJob implementation to manage Hadoop clusters inside an environment managed
by an HPC scheduler, such as PBS, SLURM or SGE, or clouds. The \pilot-Framework
is used for dispatching a bootstrap process that generates the necessary
configuration files and for starting the Hadoop processes. The specifics of the
Hadoop framework (i.\,e.\ YARN and Spark) are encapsulated in an adaptor. The
bootstrap process is then responsible for launching YARN's resource and node
manager processes respectively the Spark master and worker agents on the nodes
allocated by the \pilot-Framework. While nearly all ABDS
frameworks (e.\,g.\ MapReduce, Tez and also Spark) support YARN for resource
management, Spark provides a standalone cluster mode,
which is more efficient for dedicated resources. Thus, a special adaptor for 
Spark is provided. 

Once the cluster is setup, users can to submit applications by using the
Pilot-API's \computeunit API to start and manage application processes.
\computeunits with type Hadoop and Spark are then forwarded to the YARN
respectively Spark resource manager, which then handles the management of these
tasks. With this capability, the \pilot-Abstraction can be used to manage
highly heterogeneous workloads, e.\,g.\ bag-of-tasks, coupled tasks, MPI,
Hadoop and Spark applications, via a single interface.

\subsection{\pilotdatainmem: A Processing Engine for Machine Learning}
\label{sec:pilotinmem}

The \pilot-Abstraction provides a low-level mechanisms to manage the data
across different, possible distributed, data stores in conjunction with their
task-based computing focusing on the stage-in and out of data related to a set
of \cus. Also, it required the developer to either manually implement data
parallelism or use a higher-level framework, such as
Pilot-MapReduce~\cite{Mantha:2012:PEF:2287016.2287020}. Another constraint is
the fact that only persistent storage can be used. The usage of (distributed)
memory for caching of input or intermediate data (e.\,g.\ for iterative machine
learning) is not supported. While this disk-based model is effective for doing
many forms of large volume data processing, iterative processing, e.\,g.\ for
machine learning, requires more sophisticated ways to manage intermediate data.

\emph{\pilotdatainmem} adds in-memory capabilities to \pilotdata
and makes it available via the \pilot-API. A particular
challenge is the integration of the in-memory layer with the compute
layer, which typically requires support for a manifold set of tools and programming languages. We focus on providing a Python-based API and runtime;
many scientific applications are implemented in Python, which makes it
easy to integrate native code to achieve a good performance. 
Further, there is a lack of in-memory frameworks and tools for Python --
\pilotdatainmem is an attempt to address these limitations. A
particular gap is the ability to manage large amounts of distributed
memory - while it is fairly simple to manage memory on a single node
(using e.\,g. memory mapped files), support for distributed memory
typically requires specialized runtime and processing environments
that are not compatible with traditional Posix file APIs.

To process data in an in-memory \du, we extend the \du interface to
provide a higher-level MapReduce-based API for expressing
transformations on the data. Using a map and reduce functions,
applications can express abstract operations on data
without manually creating \cus for partitioning and processing the
data. The API utilizes a key/value pair tuples as input for the map
and reduce function. The runtime system generates the necessary
application tasks (\computeunits) and run these in parallel
considering data locality and other aspects. Users have the
possibility to control this placement using a simple, label-based
affinity model, which allows reasoning about compute and data and
provides the runtime system with hints for optimizing execution. The
system is data format agnostic and supports heterogeneous data schemes
(schema on read).

\begin{figure}[t]
    \centering
       \includegraphics[width=.47\textwidth]{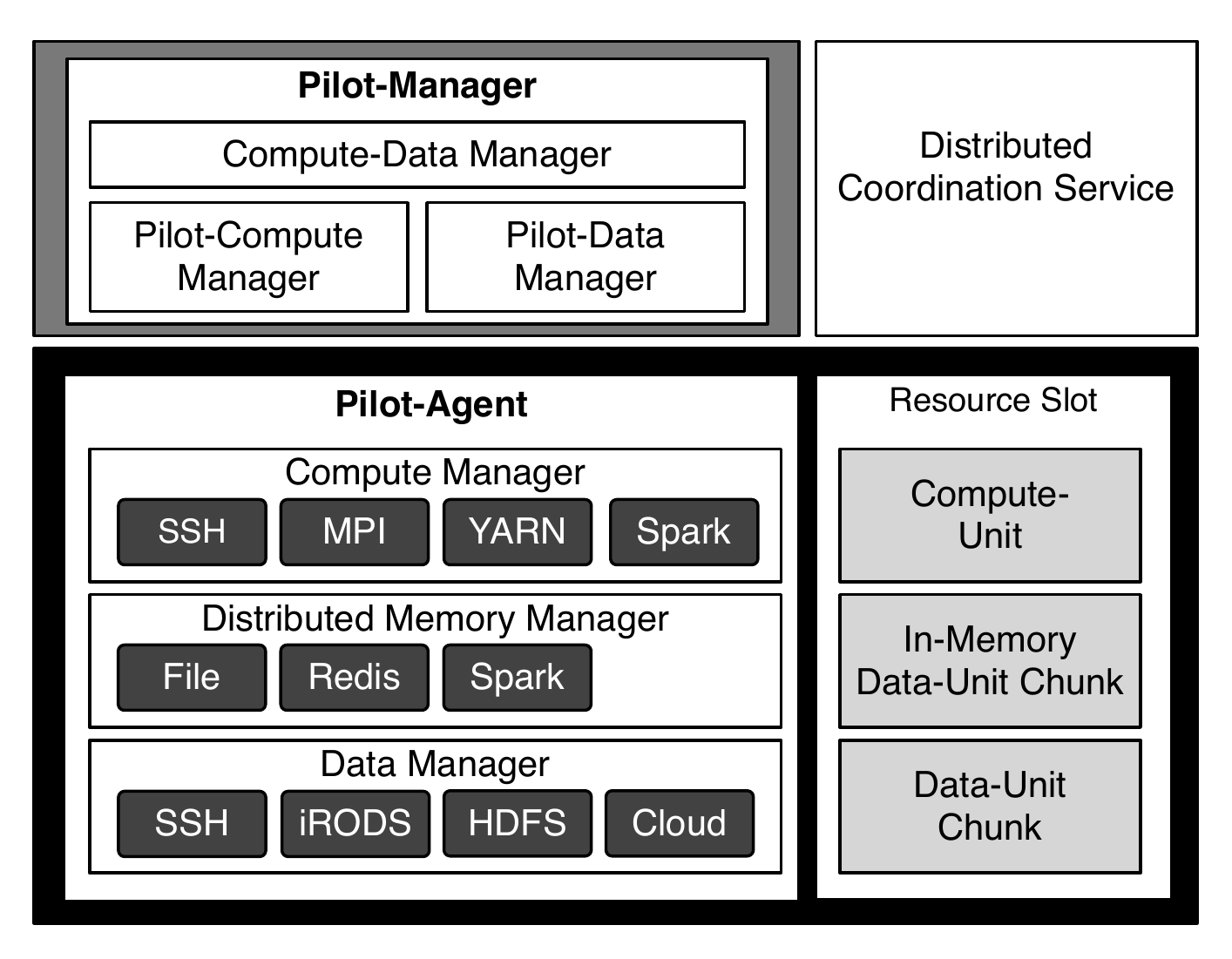}
       \caption{\pilotdatainmem Architecture: The \pilot-Manager handles 
       manages a set of queues from which the agents are pulling \cus. Data     
       management is carried out at the \pilot-Agent-level. The \pilot-Agent 
       will stage-in and out data via the Data Manager. The Distributed Memory 
       Manager handles the caching of data. \cus are executed via the Compute 
       Manager.
 }
    \label{fig:figures_pilotdata-hadoop-architecture}
\end{figure}

Figure~\ref{fig:figures_pilotdata-hadoop-architecture} illustrates the
architecture of the \pilotdatainmem framework. There are multiple levels of
coordination and decision making: The Compute-Data-Manager manages a set of
memory and disk-backed \pilotdata and \pilotcompute. Applications can submit
\computeunits (\cus) and \dataunits (\dus) to the \pilot-Manager, which exposes
control over \pilots, \cus and \dus via the \pilot-API. The
Compute-Data-Manager will assign submitted \computeunits and \dataunits to a
\pilot taking into account the current available \pilots, their utilization and
data locality. The \pilot-Agent will stage-in and out data via the data
manager. The Distributed Memory Manager handles the caching of data required
for the computation. \cus are executed via the Compute Manager. The
architecture enables late decision making at runtime: depending on the current
utilization \cus can be processed by different \pilots. Currently, the
\pilot-Manager considers both the utilization of the \pilot and data locality.
In the future, we plan to support further resource characteristics, such as the
amount of available memory (critical for memory centric computing), as well as
the characteristics of the \cu workload (providing support for co-placing \cus
or streaming data between two subsequent \cus).

An important design objective for \pilotdatainmem is extensibility and
flexibility. Thus, \pilotdata provides an adaptor mechanism to support different
in-memory backends. Currently three different backends are supported: (i)
file-based, (ii) in-memory Redis and (iii) in-memory Spark. \pilot-Hadoop can
be used to setup the necessary Spark infrastructure on a HPC resource. We
further evaluate support for Tachyon~\cite{tachyon} and the HDFS In-Memory
storage tier~\cite{hdfs-inmen}. The adaptor service interface specifies the
capabilities that need to be implemented by the in-memory backend; it consists
of functions for allocating/deallocation memory, for loading data and for
executing a map and reduce functions on the data. Depending on the backend the
processing function need to be implemented either manually, e.\,g.\ for the
file-based and Redis backend adaptor, or can be directly delegated to the 
processing engine as for Spark. The Redis and file backends use the \pilotjob 
framework for executing the \cus generated by \pilotdatainmem.

In summary, the \pilot-Abstraction and \pilotdatainmem allows applications to
seamlessly move data between different forms of storage and memory providing
the basis for the implementation of complex data workflows, e.\,g.\ for fusing
different data sources, data filtering, feature extraction and for execution
complex analytics on top of the data. As described in the following section
(section~\ref{sec:experiments}), \pilotdatainmem enables the efficient
implementation of advanced analytics algorithms allowing e.\,g.\ the efficient
storage of intermediate data in memory for iterative processing.

\section{Experiments}
\label{sec:experiments}

We proposed several extensions to the \pilot-Abstraction to accommodate Hadoop
infrastructures. In this section, we investigate the characteristics of the
Hadoop adaptor for the \pilot-Framework BigJob and \pilot-Hadoop. Further, we
evaluate the performance of Hadoop MapReduce on HPC resources and compare HDFS
and Lustre. Finally, we use KMeans to validate the suitability of
\pilotdatainmem for data analytics applications.

\subsection{Application Framework and Backend Interoperability}

In the following, we investigate the performance of the new \pilotdata adaptors
for YARN and Mesos as well as \pilot-Hadoop. For this experiments, we utilize
Mesos 0.14, Hadoop 2.6 and Spark 1.1 as well as Amazon EC2 and the XSEDE
machine Stampede.

\begin{figure}[t]
     \centering
     \includegraphics[width=.5\textwidth]{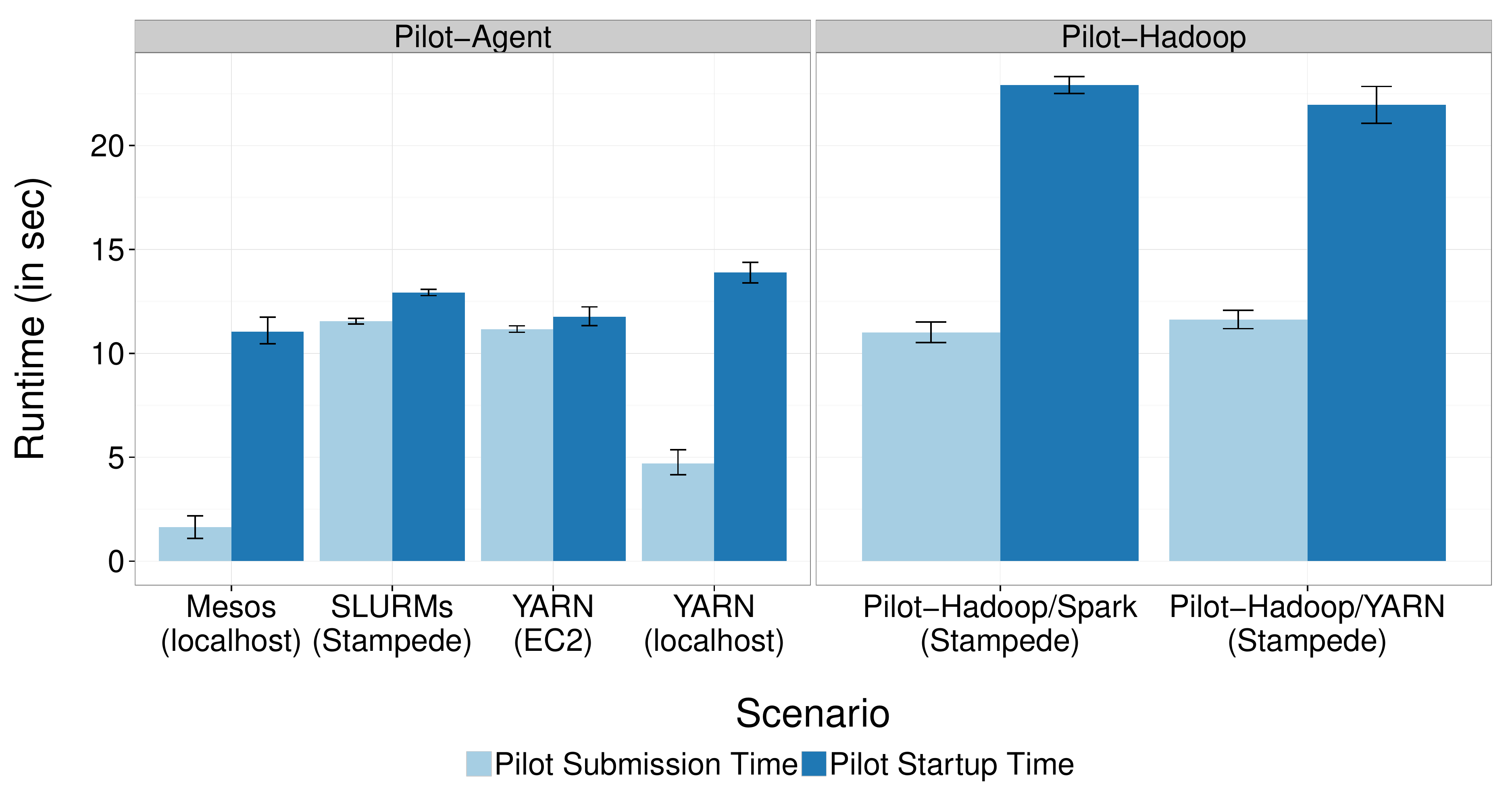}
     \caption{\textbf{Pilot-Data: Interoperable Use of Pilot-Data on 
     HPC, Cloud and Hadoop Resources:} Pilot-Data supports different
     compute infrastructure using an adaptor mechanism and SAGA (left facet).
     Further, it supports running ABDS frameworks, such as Hadoop's YARN and
     Spark, on HPC and Cloud environments (right facet).}
     \label{fig:experiments_startup_startup}
 \end{figure}

Figure~\ref{fig:experiments_startup_startup} summarizes the
results. The left part illustrates the usage of the native Pilot-Data
to manage \cus on HPC (Stampede) and on YARN and Mesos clusters (in
this case hosted on Amazon Web Services). For short-running jobs the
scheduler often represents a bottleneck, e.\,g.\ the startup of a
YARN application typically requires several seconds mainly to due the
overhead induced by the JVM startup as well as the complex startup
process; resources have to be requested in two stages: first the
application master container is allocated followed by the containers
for the actual compute tasks.

For the \pilot-Hadoop scenarios we utilize the Stampede supercomputer. In
addition to the normal overhead for starting the \pilot-Agent (see left facet
of Figure~\ref{fig:experiments_startup_startup}), some extra time is needed for
setting up a YARN respectively Spark cluster on Stampede. In both cases, we
will spawn the YARN and Spark daemons without HDFS assuming that data will be
read from the Lustre storage cluster. Both YARN and Spark show a comparable
startup time.

\subsection{Exploring Hadoop on HPC}

\pilot-Hadoop enables users to start YARN and Spark clusters on HPC resources
managed by a scheduler, such as PBS, SLURM or SGE. While the mechanics of
launching Hadoop on HPC resources are well understood, a challenge remains the
configuration of Hadoop in an optimal way taking into account the specifics of
the resource, such as the available memory, storage (flash vs. disks) etc., as 
these can vary even within a single infrastructure, such as XSEDE. 

The objective of this experiment is to explore the usage of Hadoop on two XSEDE
resources: Stampede and Gordon. On Stampede the storage space is partitioned
into home, work, scratch and archival storage. The home, work and scratch
directories are located on a Lustre filesystem; they differ with respect to
their quota, backup and purge policy. The archival storage is located on a
remote system and not directly mounted on the compute nodes. The amount of
local space is constrained to 80\,GB in the /tmp directory. Some resources
started to cater more data-intensive workloads. Gordon e.\,g.\ offers 280\,GB
flash-based local storage per node. However, since the space is transient, this
space is mainly suitable for intermediate data -- otherwise data needs to be
initially copied to this space. In addition to different storage option, we investigate the usage of the new HDFS In-Memory feature~\cite{hdfs-inmen}.

\begin{figure}[t]
    \centering
\includegraphics[width=.49\textwidth]{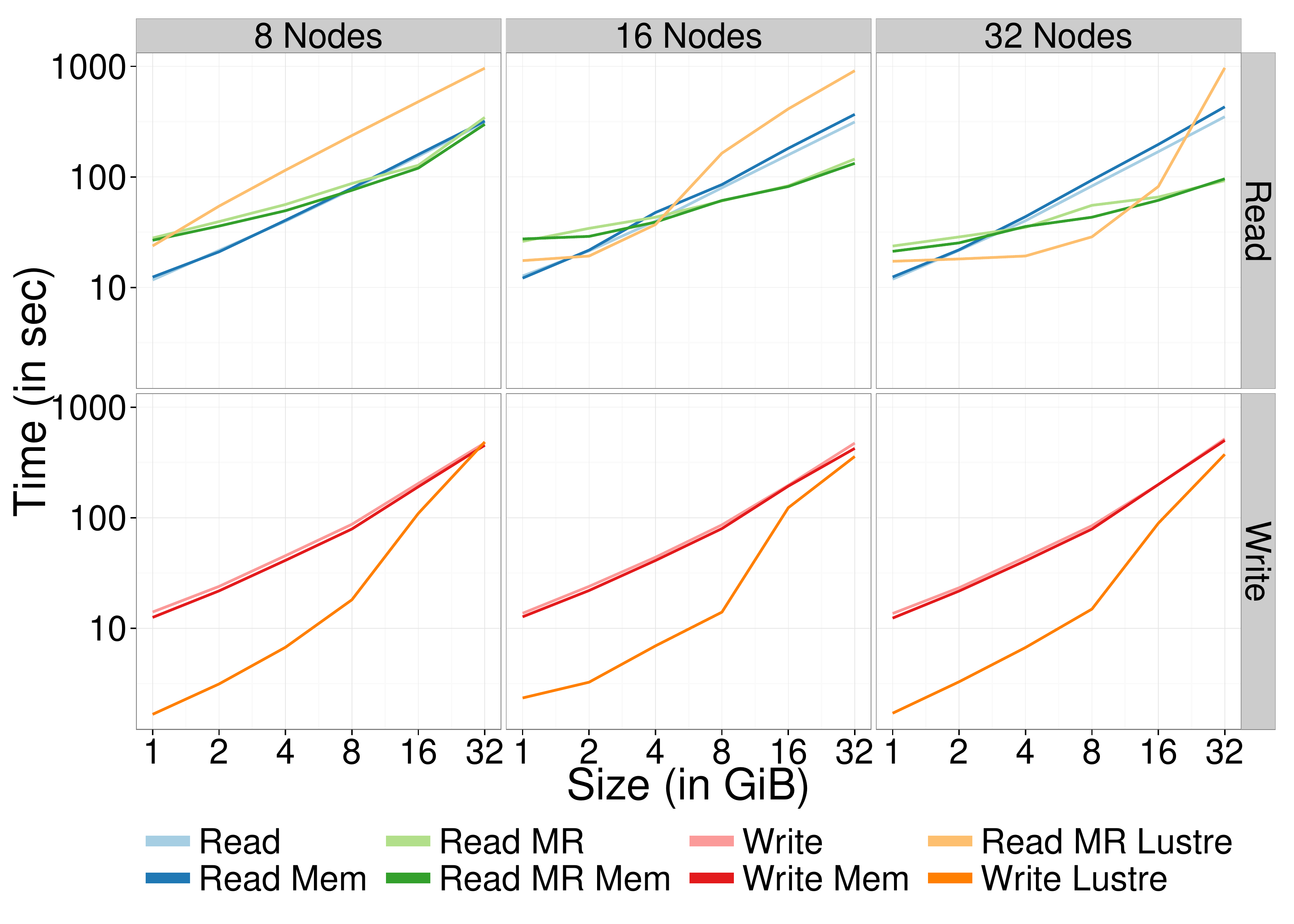}
\caption{HDFS and Lustre Performance on Stampede: The figure shows the time to 
read/write different data sizes data to/from HDFS and Lustre. Lustre performs 
well for small data sizes, while HDFS has a better performance for 
MapReduce-based parallel reads of larger data sizes on larger clusters.}
    \label{fig:experiments_iterative_hdfs_inmemory_hdfs_mr_read_write}
\end{figure}

In the first step we analyze the HDFS and in particular the HDFS in-memory
performance. For this purpose, we deploy an HDFS cluster on Stampede using
Pilot-Hadoop using up to 512 cores and 32 nodes. HDFS is configured to use the
local filesystem (/tmp) as data directory. Half of the data nodes memory is
reserved for the in-memory cache. Further, we compare the HDFS performance with
the Lustre storage available on Stampede.
Figure~\ref{fig:experiments_iterative_hdfs_inmemory_hdfs_mr_read_write}
summarizes the results of our experiments on Stampede.

HDFS 2.6 provides storage policies, which enable clients to directly store data
in the memory of the data node without the need to wait for the persistence and
replication of the data. This feature requires that the native libraries for
Hadoop are in place and that the memory parameter for HDFS data node
(\texttt{datanode.max.locked.memory}). We see a consistent minor improvement of
the write performance when using the in-memory option. However, the overall
write performance is determined by the non-parallel write to HDFS. We
investigate two kinds of read performances: (i) the read performance using a
single client that executes a get command and (ii) the parallel read using a
MapReduce job. The larger the cluster size, the better the parallelism for
MapReduce parallel reads -- as expected Hadoop scales near linear in this case.
For the in-memory case, we see no performance improvements for normal reads
(case (i)); for (ii) we see a minor benefits.

Further, we investigate the performance of Lustre. In particular for small file
sizes, Lustre performs well. Surprisingly, also the read performance using a
non parallel-I/O client was in many cases lower than in the non-parallel HDFS
client (case (i)). Obviously, for MapReduce workloads HDFS clearly outperforms
Lustre. MapReduce utilizes data locality when assigned map tasks to data file
chunk. While there is a Hadoop Lustre plugin as part of the Intel Enterprise
Lustre edition available that utilizes a similar mechanism, it is currently not
available on Stampede. Thus, by default when running Hadoop MapReduce on top of
Lustre data-locality is not considered.

\begin{figure}[t]
    \centering
\includegraphics[width=.49\textwidth]{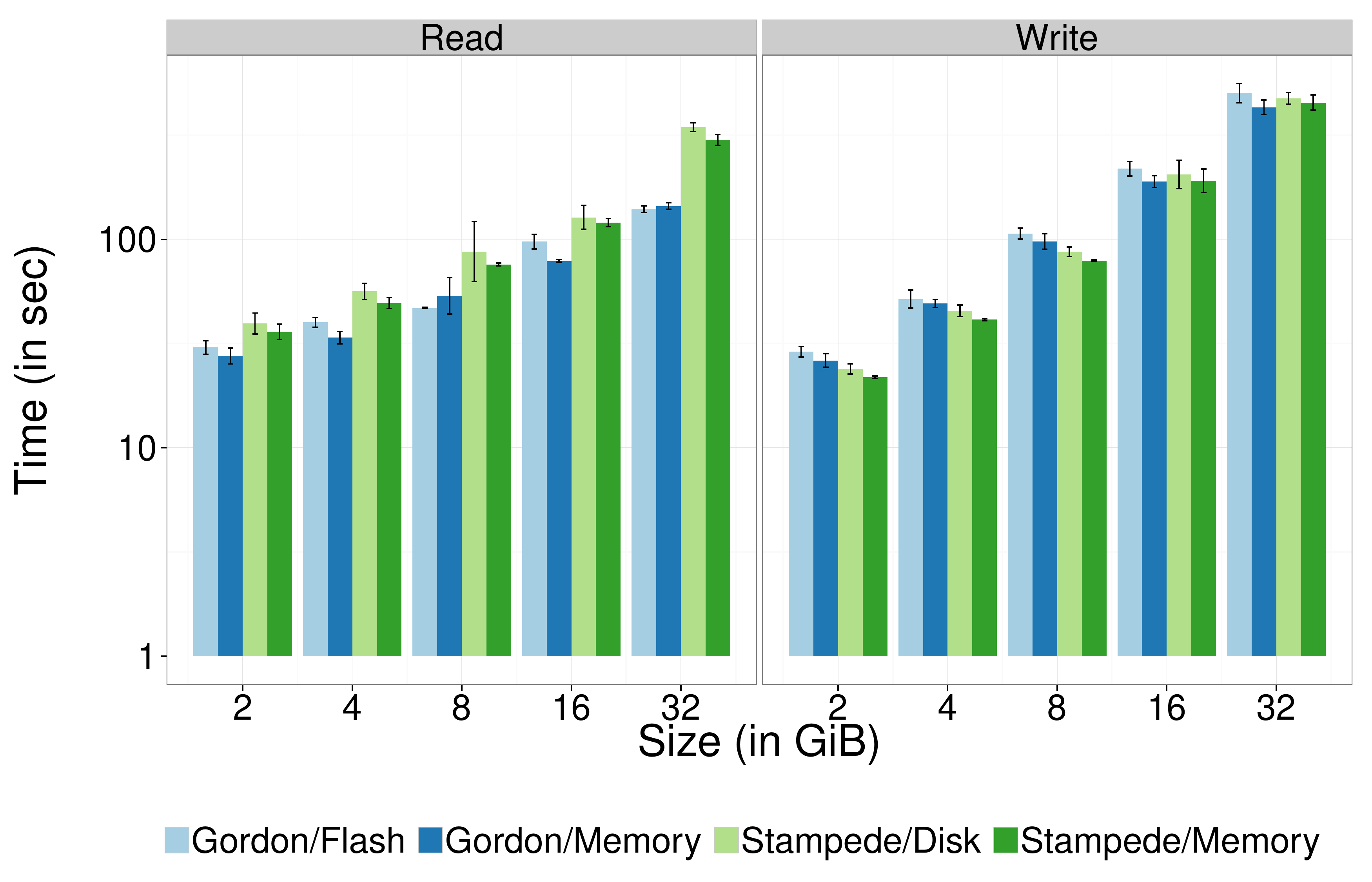}
    \caption{Gordon vs. Stampede on 8 Nodes/128 Cores: The HDFS performance on 
    Gordon is significantly better mainly due to the usage of local flash 
    drives compared to disks. The speedup between flash storage and memory on 
    Gordon is as expected lower than between disk and memory on Stampede.}
\label{fig:experiments_iterative_hdfs_inmemory_stampede_vs_gordon_hdfs_8nodes}
\end{figure}

Another concern is the performance of different environments; in the following
we compare the performance of Stampede and Gordon.
Figure~\ref{fig:experiments_iterative_hdfs_inmemory_stampede_vs_gordon_hdfs_8nodes} shows the performance of both environments. Gordon clearly shows a better
performance for HDFS mainly due to the local flash storage and more memory in
the machines. The performance improvement for the in-memory option is in
average 9\,\% in comparison to Stampede where the speedup for in-memory is in
average 14\,\%.

HDFS heterogeneous storage supports provides a uniform interface for reasoning
about storage across a single namespace, which allows a simplification of
existing data flows. However, not all YARN/HDFS based tools optimally utilize
these capabilities. MapReduce e.g. does not utilize the in-memory HDFS
features. In the future, it can be expected that frameworks such as Spark will
make use of these capabilities.

\subsection{Advanced Analytics: KMeans}

KMeans is a classical example of an advanced analytics algorithms used for
clustering data that shares characteristics with a broad set of other
analytics algorithms. The algorithms requires multiple iterations on
the data -- in each iterations a new candidate location for the
cluster centers.

In the following, we use our \pilotdata-based KMeans implementation
to evaluate different \pilotdatainmem backends. The
computational kernel, which is called iteratively takes two parameters: (i) a
set of multi-dimensional set of vectors to be clustered, remains constant
through all the iterations, and (ii) a set of centroid vectors. The centroids
vector changes each iteration. For the \pilot-based Redis backend, we utilize
one \pilotcompute managing up to 384\,cores on Stampede for running \cus on the
\dataunit stored in a \pilotdatainmem. The file backend is included for
reference purposes. For the Spark scenario, we utilize \pilot-Spark to setup a
Spark cluster on Stampede -- an XSEDE leadership machine, which is then used to
manage both the data and compute inside of Spark (via the \pilot-API). We use
Redis 2.8 and Spark 1.1. The experiment demonstrates interoperability in two
dimensions: first, it shows how ABDS frameworks can be run on HPC
infrastructures, second it demonstrates the versatility of the
\pilot-Abstraction for implementing data analytics algorithms on top of
different in-memory runtimes in an infrastructure agnostic way.

\begin{figure}[t]
    \centering        \includegraphics[width=.49\textwidth]{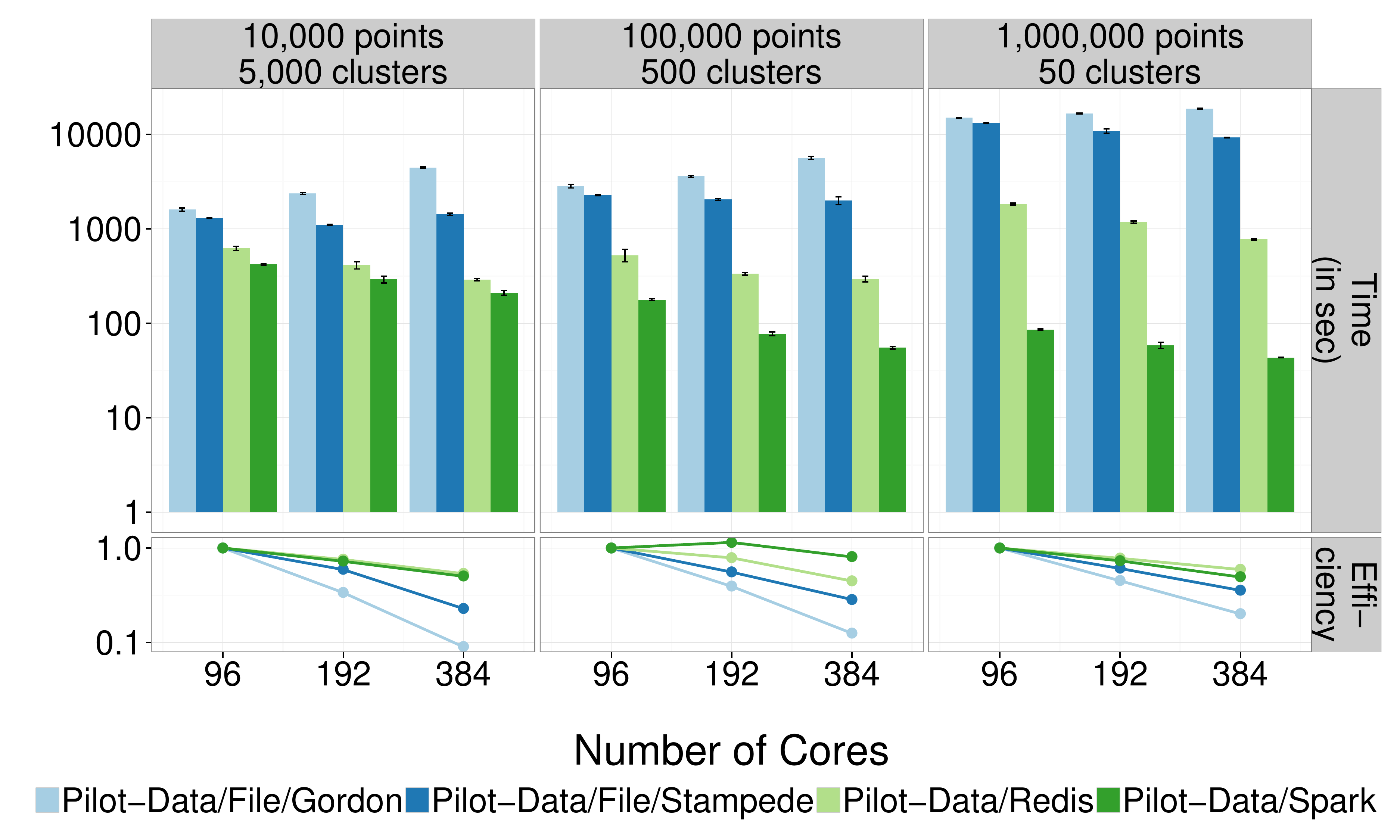}
    \caption{\textbf{\pilotdatainmem KMeans:} We compare three Pilot-Data 
    backends: (i) file-based and two in-memory-based: (ii) Redis and (iii) 
    Spark. The Redis and Spark 
    scenario are both executed on Stampede. KMeans on \pilotdata/Spark has the 
    best performance with a speedup of up to 212 compared to the 
    \pilotdata/File on Stampede. Also the scale-out efficiency of Spark is in most cases better than for Redis.}
    \label{fig:experiments_kmeans_kmeans}
\end{figure}

Figure~\ref{fig:experiments_kmeans_kmeans} shows the results of the
experiments. We investigated three different K-Means scenarios: (i)
1,000,000 points and 50 clusters, (ii) 100,000 points and 500
clusters and (iii) 10,000 points and 5,000 clusters. Each K-Means
iteration comprises of two phases that naturally map to the MapReduce
programming model of \pilotdatainmem: in the map phase the closest
centroid for each point is computed; in the reduce phase the new
centroids are computed as the aver- age of all points assigned to this
centroid. While the computational complexity is defined by the number
of points $\times$ number of clusters (and thereby a constant in the
aforementioned scenarios), the amount of data that needs to be
exchanged during the shuffle phase increases gradually from scenario
(i) to (iii), with the number of points.

The performance of \pilot-KMeans improves significantly using the memory-based
runtimes for the both the data points and intermediate cluster locations. The
Redis backend achieves only a speedup of up to 11, which is significantly lower
that the speedup of up to 212 achieved with Spark. Also, the scale-out
efficiency for Spark is better than for Redis in most cases. This is mainly
caused by the fact that we utilize a non-distributed Redis server. In the
future, we will evaluate a Redis cluster setup. Both in-memory backends
scale more efficient than the file-backend.

\section{Conclusion and Future Work}
\label{sec:conclusion}

As the needs and sophistication of scientific applications increase,
the infrastructures to support these applications are becoming
increasingly heterogeneous and trying to accommodate an increasing
number and diversity of different workloads. In response, different
infrastructures have emerged: HPC for tightly-coupled applications,
HTC for task-level parallelism, clouds for elastic workloads and
Hadoop for data-intensive workloads. The deluge of tools, abstractions and 
infrastructures lead to complex landscape of point solutions characterized by a
tight coupling of the components and limited interoperability.

In some ways HPC and Hadoop environments are converging: increasingly parallel
and in-memory computing concepts have emerged in the ABDS environment, e.\,g.\
MLLib/Spark utilizes fine-grained parallelism for implementing linear algebra
operations for advanced analytics. Although the introduction of ABDS concepts
and frameworks have begun, their uptake remains stymied by multiple reasons,
one of which is related to finding satisfactory and scalable resource
management techniques usable for ABDS frameworks on HPC infrastructure.

Further, there is a need to map different aspects and stages of a data-intensive
workflow to the appropriate infrastructure. Choosing the right infrastructure
for an application however, is a difficult task: data-intensive workflows
typically comprise of multiple stages with different compute and IO
requirements. For example, while data filtering and processing is best done
with Hadoop (using e.\,g. the MapReduce abstraction), the compute-bound parts of
that workflow are best supported by HPC environments.

The \pilot-Abstraction and the \pilotdata implementation enable applications to
utilize and explore various paths of running Hadoop on HPC (vice versa)
supporting end-to-end data workflows as well as specific steps, such as
iterative machine learning. In this paper, we demonstrated the usage of
\pilot-Abstraction as a common, interoperable framework for ABDS and HPC,
supporting a diverse set of workloads, e.\,g.\ both I/O bound data preparations
tasks (using MapReduce) as well as memory-bound analytics tasks (such as
KMeans). Using the \pilot-Abstraction, applications can combine HPC and ABDS
frameworks either by running HPC applications inside YARN or by
deploying the YARN resource manager on HPC resources using \pilot-Hadoop. Using
these capabilities, applications can compose complex data workflows utilizing a
diverse set of ABDS and HPC frameworks to enable scalable data ingest, feature
engineering \& extractions and analytics stages. Each of these steps has its
own I/O, memory and CPU characteristics. Providing both a unifying and powerful
abstraction that enables all parts of such a data pipeline to co-exist is
critical.

\pilotdatainmem provides a unified access to distributed memory that is
assigned to a \pilotcompute and shared across a set of tasks. The framework
using a pluggable adaptor mechanism to support different in-memory backends,
e.\,g.\ based on Redis and Spark. We demonstrated the effectiveness of the
abstraction for iterative analytics applications using KMeans as example.
\pilotdatainmem provides a significantly improved performance with speedups of
up to 212x compared to the file-based \pilotdata backend.

In the future, we plan to extend the \pilot-Abstraction on
heterogeneous infrastructures using an enhanced set of data-intensive
applications. To support further use cases, we will evaluate support
for further operations, e.\,g.\ to execute collectives on
data. Further, we will work on a higher-level API designed for data
analytics applications and libraries that supports the expression and
execution of data pipelines. Also, infrastructures are evolving --
container-based virtualization (based on Docker~\cite{docker}) is
increasingly used in cloud environments and also supported by
YARN. Support for these emerging infrastructures is being added to the
\pilot-Abstraction.

\noindent\textbf{Acknowledgement} This work is funded by NSF
Department of Energy Award (ASCR) DE-FG02-12ER26115, NSF CAREER
ACI-1253644 and NSF 1443054. This work has also been made possible
thanks to computer resources provided by XRAC award TG-MCB090174 and
an Amazon Computing Award to SJ. We thank Professor Geoffrey Fox for
useful discussions.


\end{document}